\documentclass[12pt]{article}
\usepackage[margin=1in]{geometry}

\usepackage[utf8]{inputenc}
\usepackage[english]{babel}

\usepackage{outlines}
\usepackage{hyperref}
\usepackage{csquotes}
\usepackage{amsmath}
\usepackage{amssymb}
\usepackage{graphicx}
\usepackage{xcolor}
\usepackage{chngpage}
\usepackage{float}
\usepackage{caption}
\usepackage{subcaption}
\usepackage{times}
\usepackage[super]{cite}
\usepackage{setspace}
\usepackage{authblk}
\usepackage{url}

\title{Finite sample performance of optimal treatment rule estimators with right-censored outcomes}
\author[1]{Michael Jetsupphasuk\thanks{Corresponding author. Email: jetsupphasuk@unc.edu}}
\author[1]{Michael G. Hudgens}
\author[2]{Jessie K. Edwards}
\author[2]{Stephen R. Cole}

\affil[1]{Department of Biostatistics, University of North Carolina at Chapel Hill}
\affil[2]{Department of Epidemiology, University of North Carolina at Chapel Hill}

\begin{document}
	
	\maketitle

	\begin{abstract}
		Patient care may be improved by recommending treatments based on patient characteristics when there is treatment effect heterogeneity. Recently, there has been a great deal of attention focused on the estimation of optimal treatment rules that maximize expected outcomes. However, there has been comparatively less attention given to settings where the outcome is right-censored, especially with regard to the practical use of estimators. In this study, simulations were undertaken to assess the finite-sample performance of estimators for optimal treatment rules and estimators for the expected outcome under treatment rules. The simulations were motivated by the common setting in biomedical and public health research where the data is observational, the outcome is a survival time subject to right-censoring, and there is interest in estimating baseline treatment decisions to maximize survival probability. A variety of outcome regression and direct search estimation methods were compared for optimal treatment rule estimation across a range of simulation scenarios. Methods that flexibly model the outcome performed comparatively well, including in settings where the treatment rule was non-linear. R code to reproduce this study's results are available on Github.
	\end{abstract}

	\section{Introduction}
	
	Decision support in precision medicine involves leveraging treatment effect heterogeneity to estimate treatment rules which map individual-level covariates to treatment recommendations that maximize an expected outcome. Under sufficient treatment effect heterogeneity, treatment rules based on individual characteristics can improve patient outcomes relative to recommending a single treatment for all patients. Though medical practitioners individualize care on a regular basis, precision medicine differs by providing a general decision-making tool that can be tested, evaluated, and scaled. In contrast, individualization by physicians depends heavily on the skill of the practitioner and is thus difficult to scale beyond the physicians' patients. Both within and outside medicine, interest in treatment effect heterogeneity is widespread and growing, as represented by the explosion of research in this area across a variety of fields including statistics, medicine, epidemiology, economics, and computer science. 
	
	For outcomes without right-censoring, practical considerations for estimating optimal treatment rules and the corresponding value function (i.e., the expected outcome under a treatment rule) have been discussed in recent literature. Power analyses and sample size calculations for sequential randomized trials designed to determine optimal treatment rules have been developed \cite{artman_power_2018,turchetta_bayesian_2022,rose_sample_2019}. Empirical studies have also been conducted comparing different estimators of the value function for data with outcomes that were not censored \cite{montoya_estimators_2023}.
	
	The use of optimal treatment rule estimation in medicine and public health differs from other application areas because there is often interest in optimizing survival outcomes such as $t$-year survival probability or restricted mean survival time (RMST) when survival times are right-censored. Recently, a variety of methods have been developed to estimate both optimal treatment rules and the expected survival outcome under a treatment rule when right-censoring is present \cite{bai_optimal_2017,cui_estimating_2023,goldberg_q-learning_2012,jiang_doubly_2017,leete_balanced_2019,sparapani_nonparametric_2016,zhao_doubly_2015}. The finite sample properties of several of these estimators are compared in this paper. 
	
	This research is motivated by a recent observational study analysis of patients living with HIV \cite{jetsupphasuk_optimizing_2023} that sought to estimate an optimal treatment rule to maximize 5-year survival probability of the composite outcome of experiencing an AIDS-defining illness, a serious non-AIDS event, or all-cause death. The data from the study included nearly 16,000 patients from the North American AIDS Cohort Collaboration on Research and Design (NA-ACCORD) \cite{gange_cohort_2007} collected between 2009 and 2016. Two classes of drugs were considered at the time of treatment initiation: an efavirenz (EFV)-based drug regimen and an integrase strand transfer inhibitor (InSTI)-based regimen. The 5-year survival times of about 90 percent of patients were right-censored. This study represents a common setting in biomedical and public health research, characterized by: a single-stage binary treatment, observational time-to-event data with right-censoring, and survival probability as the expected outcome of interest. Using the attributes of this study as motivation, the present paper evaluates the finite sample performance of a variety of estimators for the optimal treatment rule and value function over several simulation scenarios. 
	
	The remainder of this paper is organized as follows. In Section 2, the notation, parameters of interest, and estimators are defined. In Section 3, the data generating process and performance metrics are detailed. In Section 4, the simulation findings are summarized. Section 5 concludes with a discussion.
	
	\section{Methods}
	\subsection{Notation and assumptions}
	Consider an observational study of $n$ individuals where each individual indexed $i=1,\ldots,n$ has survival time $T_i$, right-censoring time $C_i$, and binary treatment $A_i$ equal to 0 or 1. Only one of the survival time or right-censoring time may be observed, such that the observed data includes $Z_i=\min(T_i,C_i)$ and the event indicator $\Delta_i=I(T_i\leq C_i)$ where $I(\cdot)$ denotes the indicator function. Let $X_i$ be the $p$-dimensional vector of baseline covariates. Thus, the quadruple $(Z_i,\Delta_i,X_i,A_i)$ is observed for each individual $i=1,\dots,n$. 
	
	The methods described below rely on the following standard causal assumptions for right-censored outcomes. Let $T_i^*(a)$ denote the potential survival times had, possibly counter to fact, individual $i$ received treatment $a\in\{0,1\}$. Survival times are assumed to follow the causal consistency assumption, i.e., $T=T^*(1)A+T^*(0)(1-A)$. Additionally, positivity of treatment selection and censoring, or $0<P(A=1|X)<1$ and $P(C>T|A,X)>0$, are assumed. Censoring is assumed to be non-informative for potential survival, conditional on covariates, $C\perp T^*(a)|X,A$ for $a=0,1$ where $\perp$ denotes statistical independence. Lastly, no unmeasured confounding, $A\perp T^*(a)|X$ for $a=0,1$, is assumed. 
	
	\subsection{Value function and treatment rule}
	
	A treatment rule $d$ is a function that maps covariates $X$ to a treatment $A \in \{0,1\}$. The value function $V(d)$ is defined in general to be the expected outcome had, potentially counter to fact, all patients received the treatment specified by the treatment rule $d$. Assuming the outcome of interest is coded such that higher values are preferable, the optimal treatment rule $d^\mathrm{opt}$ is defined to be the treatment rule that maximizes $V(d)$. In general, the outcome of interest will be a function of the survival time. Here, the outcome of interest is survival at some time point $\tau$, i.e., $I(T^*(d)>\tau)$, where $T^*(d)=d(X)T^*(1)+(1-d(X))T^*(0)$. Thus, the value function is the probability of survival at time $\tau$, i.e., $V(d)= \mathbb{E} [I(T^*(d)>\tau)]$, under treatment rule $d$. 
	
	Static rules assign the same treatment to all patients, regardless of individual characteristics. In particular, the treatment rules $d(X)=0$ and $d(X)=1$ are referred to as static rules $0$ and $1$, respectively. The optimal treatment rule $d^\mathrm{opt}$ differs from a static rule if there is sufficient treatment effect heterogeneity with respect to the covariates $X$. The regret of a treatment rule $d$ is defined as $R(d)=V(d^\mathrm{opt}) - V(d)$, which represents the increase in the value function afforded by the optimal treatment rule $d^\mathrm{opt}$ relative to treatment rule $d$. 
 
	For a given treatment rule $d$, three consistent inverse probability of treatment and censoring estimators for the value function were considered: a Kaplan-Meier (IPW-KM) estimator, a Horvitz-Thompson (IPW-HT) estimator, and a Hajek (IPW-Hajek) estimator. The IPW-KM estimator \cite{jiang_estimation_2017} is given by 
	\begin{equation*}
		\widehat{V}_1(d) = \prod_{s\leq \tau} \left\{1-\frac{\sum_{i=1}^n \widehat{w}_i(s;d)dN_i(s)}{\sum_{i=1}^n \widehat{w}_i(s;d)Y_i(s)}\right\}
	\end{equation*}
	where $N_i(s)=I(Z_i\leq s,\Delta_i=1)$ is the observed event counting process at time $s$, $Y_i(s)=I(Z_i\geq s)$ is the at-risk process at time $s$, and the estimated weights are given by
	\begin{equation*}
		\widehat{w}_i(s;d)=\frac{I(A_i=d(X_i))}{[\widehat{\pi}(X_i)A_i+(1-\widehat{\pi}(X_i))(1-A_i)]\widehat{P}(C_i\geq s|X_i,A_i)}
	\end{equation*}
	where $\widehat{P}(C_i\geq s|X_i,A_i)$ is an estimator of $P(C_i\geq s|X_i,A_i)$, the probability of being uncensored at time $s$, and $\widehat{\pi}(X_i)$ is an estimator of the propensity score $\pi(X_i)=P(A_i=1|X_i)$ \cite{jiang_estimation_2017}. Expressing the value function as 
    \begin{equation*}
        V(d)=E\left[\frac{\Delta_i I(T_i>\tau)I(A_i=d(X_i))}{P(C_i\geq \tau|X_i,A_i)[\pi(X_i)A_i+(1-\pi(X_i))(1-A_i)]} \right]
    \end{equation*} motivates the form of the IPW-HT estimator \cite{zhao_doubly_2015}:
	\begin{equation*}
		\widehat{V}_2(d) = \frac{1}{n} \sum_{i=1}^n \Delta_i I(T_i>\tau) \widehat{w}_i(\tau;d)
	\end{equation*}
	The IPW-Hajek estimator tends to have lower variance than the IPW-HT estimator and is given by:
	\begin{equation*}
		\widehat{V}_3(d) = \frac{n}{\sum_{i=1}^n \frac{I(A_i=d(X_i))}{\widehat{\pi}(X_i)}} \widehat{V}_2(d)
	\end{equation*}
	The above value estimators are consistent estimators of $V(d)$ if the nuisance functions ${P}(C_i\geq s|X_i,A_i)$ and ${\pi}(X_i)$ are consistently estimated by their respective estimators.

	\subsection{Estimators for the optimal treatment rule} \label{sec:sample_split}
	
	Estimators for the optimal treatment rule can be broadly divided into two classes: outcome regression methods and direct search methods. Outcome regression methods involve expressing the optimal treatment rule as $d^{\mathrm{opt}}(X) = I(P(T > \tau | A=1, X) > P(T > \tau | A=0, X))$. Regression modeling can then be used to estimate $P(T > \tau | A, X)$ leading to a plug-in estimator of $d^{\mathrm{opt}}(X)$. The choice of model restricts the space of possible estimated treatment rules. For example, if a linear regression model is chosen to estimate $P(T > \tau | A, X)$, then the estimated optimal treatment rule will also be linear, regardless of whether the true optimal treatment rule is linear. The regression methods considered in the simulations described below included penalized Cox regression (with ridge, Lasso, and elastic net penalties) \cite{tibshirani_regression_1996,zou_regularization_2005}, causal survival forests \cite{cui_estimating_2023}, and Bayesian additive regression trees (BART) \cite{sparapani_nonparametric_2016}. 
 
    Direct search involves using numerical or machine learning methods to estimate an optimal treatment rule by directly optimizing the value function, usually restricting the class of possible treatment rules to a pre-specified space. The direct search methods compared in the simulation study included the use of a genetic algorithm and imputation combined with support vector machines. The genetic algorithm method entailed numerically searching over a specified space of parametric treatment rules to optimize the IPW-KM estimator and variations where the IPW-KM estimator is augmented by an outcome regression model, the treatment rule is smoothed, or both \cite{jiang_estimation_2017}. The augmented IPW-KM estimator is given by
    \small{
    \begin{equation*}
        \widehat{V}_{1,aug}(d) = \prod_{s\leq \tau} \left\{1-\frac{\sum_{i=1}^n [ \widehat{w}_i(s;d)dN_i(s) + (1- \widehat{w}_i(s;d))\widehat{P}(T > s| d(X_i), X_i) \widehat{P}(C_i\geq \tau|X_i,A_i) \widehat{h}_T(s| d(X_i),X_i) ]}{\sum_{i=1}^n [\widehat{w}_i(s;d)Y_i(s) + (1- \widehat{w}_i(s;d))\widehat{P}(T > s| d(X_i), X_i) \widehat{P}(C_i\geq \tau|X_i,A_i)]} \right\}
    \end{equation*}
    }
    where $\widehat{h}_T(s| d(X_i),X_i)$ is the estimated conditional hazard function at time $s$. For either the augmented or non-augmented IPW-KM estimators, the treatment rule may be smoothed which may improve estimation. For example, one can specify the estimated treatment rule to be linear, i.e., $X_i^T \theta$, where $\theta$ is a $p \times 1$ vector of coefficients. Then, $d(X_i) = I(X_i^T \theta > 0)$ which may be replaced with the smoothed $d_{smooth}(X_i) = \Phi(\frac{1}{h}X_i^T \theta)$ where $\Phi$ is the standard normal distribution function and $h$ is a parameter that controls smoothness. The genetic algorithm then searches over values of $\theta$ until a stopping criterion is reached. 
    
    Outcome weighted learning (OWL) \cite{zhao_doubly_2015} searches over a specified parameterization of the treatment rule to optimize the IPW-HT estimator while residual weighted learning (RWL) \cite{zhou_residual_2017} augments the estimator with an outcome regression model. Recursively imputed survival trees \cite{zhu_recursively_2012} were used to impute right-censored outcomes for use in the outcome and residual weighted learning estimators \cite{cui_tree_2017}. In both cases, the optimization problem is converted to a weighted classification problem where supervised machine learning tools such as support vector machines can be used. In general, finding the $d$ that optimizes the IPW-HT estimator can be posed as a weighted classification problem by noticing that maximizing $\widehat{V}_2(d)$ with respect to $d$ is equivalent to finding the $d$ which minimizes 
    \begin{equation*}
        \sum_{i=1}^{n} \frac{\Delta_i I(T_i>\tau)I(A_i \neq d(X_i))}{\widehat{P}(C_i\geq \tau|X_i,A_i)[\widehat{\pi}(X_i)A_i+(1-\widehat{\pi}(X_i))(1-A_i)]}
    \end{equation*}
    or, in other words, the optimization attempts to minimize the mis-classification of $A_i$ given $X_i$ subject to mis-classification error weights
    \begin{equation*}
        \frac{\Delta_i I(T_i>\tau)}{\widehat{P}(C_i\geq \tau|X_i,A_i)[\widehat{\pi}(X_i)A_i+(1-\widehat{\pi}(X_i))(1-A_i)]}
    \end{equation*}
    When support vector machines are used for optimization, a kernel function that parameterizes the treatment rule must be specified. In the simulation studies below, linear and radial kernels were used. 

    Overfitting may be a concern when the estimated optimal treatment rule $\widehat{d}^{\text{opt}}$ is obtained by optimizing $\widehat{V}(d)$, potentially resulting in $\widehat{V}(\widehat{d}^{\text{opt}})$ overestimating $V(\widehat{d}^{\text{opt}})$. A similar phenomenon has been shown in other settings \cite{athey_recursive_2016,jiang_precision_2021}. Commonly, sample splitting is used to ameliorate this problem. The simplest form of sample splitting entails randomly splitting the data into two sets where one set is used for estimation of the optimal treatment rule and the other for estimation of the value function. However, such sample splitting is inefficient since only a subset of the data is used for estimation of each quantity. Instead, it is common to employ a sample splitting scheme similar to $K$-fold cross-validation or the delete-$d$ jackknife. 
	
	Sample split versions of the above estimators were constructed as follows. The data indexed by $i=1,\dots,n$ were split into $K \geq 2$ partitions of approximately equal size $CV_1, \dots, CV_K$. For, $k \in \{1,\dots,K\}$, the estimated treatment rule $\widehat{d}^{(k)}$ was computed using partitions $CV_1, \dots, CV_{k-1}, CV_{k+1}, \dots, CV_K$. Then, the sample split IPW-KM estimator of survival probability was computed by
	
	\begin{equation*}
		\widehat{V}_1(\widehat{d}^{CV})=\prod_{s\leq\tau}\left\{1-\frac{\sum_{i=1}^n\widehat{w}_i(s;\widehat{d}^{CV})dN_i(s)}{\sum_{i=1}^n\widehat{w}_i(s;\widehat{d}^{CV})Y_i(s)}\right\}
	\end{equation*}
	where $\widehat{d}^{CV}(X_i)=\sum_{k=1}^K\widehat{d}^{(k)}(X_i)I(i\in CV_k)$. The sample split versions of the IPW-HT and IPW-Hajek value function estimators were constructed analogously. 

    The simulation studies below also evaluated an ensemble estimator which was computed by estimating the treatment rule using multiple methods and then selecting the method with the highest value estimate. Specifically, this ``max ensemble estimator" was computed as 
    \begin{equation*}
         \widehat{d}^{CV}_{\max} = \arg \max_{\widehat{d}^{CV}_l} \{ \widehat{V}_1(\widehat{d}^{CV}_1), \dots, \widehat{V}_1(\widehat{d}^{CV}_L) \}
    \end{equation*}
    where $\widehat{d}^{CV}_l$ is the optimal treatment rule estimated using method $l=1,\dots,L$. 
    
    Each optimal treatment rule estimator, including the max ensemble estimator, was also computed using the whole sample (i.e., the sample splitting procedure was not used). In particular, let $\widehat{d}^{\text{whole}}_1, \dots, \widehat{d}^{\text{whole}}_L$ be estimators of the optimal treatment rule using the whole data and method $l=1,\dots,L$, and let $\widehat{d}^{\text{whole}}_{\max} = \arg \max_{\widehat{d}^{\text{whole}}_l} \{ \widehat{V}_1^(\widehat{d}^{\text{whole}}_1), \dots, \widehat{V}_1^(\widehat{d}^{\text{whole}}_L) \}$. In this study, the optimal treatment rule estimators $\widehat{d}^{CV}_1, \dots, \widehat{d}^{CV}_L,$ $\widehat{d}^{CV}_{\max}, \widehat{d}^{\text{whole}}_1, \dots, \widehat{d}^{\text{whole}}_L,$ and $\widehat{d}^{\text{whole}}_{\max}$ were compared where $l=1,\dots,L$ indexes the 12 optimal treatment rule estimators discussed above, i.e., Cox regression (with ridge, Lasso, and elastic net penalties), causal survival forests, Bayesian additive regression trees, OWL (with linear and radial kernels), RWL, and the genetic algorithm method (using non-augmented and non-smoothed, augmented, smoothed, and augmented and smoothed versions of the objective function).

	\section{Simulation design}
	The design of the simulations was motivated, in part, by the previously discussed observational study using data from the NA-ACCORD \cite{jetsupphasuk_optimizing_2023}. Each simulated dataset contained $n=$ 2,500 observations and the number of observed survival times in each simulation was approximately the same as in the NA-ACCORD, about 1,500. Survival times were generated to mimic the range and distribution of the NA-ACCORD data, where the time point of interest was 60 months and the dataset contained survival times measured monthly. Accordingly, in all simulations $\tau=60$ and observed survival times were coarsened to integer values. The end-of-study was defined to be time 61 so that the quadruple $(Z,\Delta,X,A)$ was observed where $Z=\lceil\min(T,C,61)\rceil$ is the coarsened observed time, $\lceil.\rceil$ is the ceiling function, and $\Delta=I(\lceil T\rceil\leq\lceil C \rceil, T \leq 61)$ is the event indicator. For each scenario, 1,000 data sets were simulated.
	
	\subsection{Covariates and treatment}
	Covariates were also generated to mimic the data from the NA-ACCORD. Nineteen covariates and a binary treatment indicator were generated to approximately match the marginal distributions of the corresponding covariates in the NA-ACCORD. See Table \ref{tab:tab1} for a comparison between the NA-ACCORD covariates and the simulated covariates. The same data generating process was used to simulate covariates for all scenarios.
	
	An unobserved latent variable was used to generate covariates. There were two latent groups of individuals, a group that was generally healthier on average and a group that was less healthy. Those in the unhealthy group had a higher probability of having a risky health behavior or adverse health condition as encoded by the binary covariates $X_{i,1}, \dots, X_{i,12}$. For notational simplicity, the $i$ subscript is omitted from the remainder of this section. The indicator for membership in the unhealthy group was $U \sim \operatorname{Bernoulli}(0.2)$, and covariates were generated as $X_{j} | U \sim \operatorname{Bernoulli}(l(p_{j}^*))$ for $j=1,\dots,12$, where $p_{j}^*=l^{-1}(p_j)+1.25(I(U=1)l(4) - I(U=0)l(-0.25))$ and $l(\cdot) = \operatorname{logit}^{-1}(\cdot)$. The parameter $p_j$ was chosen to equal the observed proportion for the corresponding covariate $X_{j}$ in the NA-ACCORD. Covariates $X_{13},\dots,X_{15}$ corresponded to binary demographic data and were generated independently: $X_{j} \sim \operatorname{Bernoulli}(p_j)$, $j=13,14,15$.
	
	Continuous covariates were generated from the normal distribution. The covariate $X_{16}$ corresponded to (natural) log age and was generated $X_{16} \sim N(\log(m_{16}), s_{16}^2)$ where $m_{16}$ was the median log age in the NA-ACCORD and $s_j$ was chosen so that the simulated interquartile range (IQR) was similar to the observed IQR in the NA-ACCORD. The other three continuous covariates $X_{17}$, $X_{18}$, $X_{19}$ corresponded to BMI, CD4 T-cell count, and HIV viral load, respectively. These were generated from a multivariate normal distribution,
	\[\begin{pmatrix} X_{17} \\ X_{18} \\ X_{19} \end{pmatrix} \sim N_3 \left( \begin{pmatrix} \log(m_{17}^*) \\ \log(m_{18}^*) \\ \log_{10}(m_{19}^*) \end{pmatrix}, \begin{pmatrix} 0.04 & 0.02 & -0.04 \\ 0.02 & 0.25 & -0.2 \\ -0.04 & -0.2 & 1 \end{pmatrix}\ \right)\]
	where $m_j^* = m_j + I(U = 0) c_{1,j} + I(U = 1) c_{2,j}$ for $j=17,18,19$, and $m_j$ was the observed median in the covariate corresponding to the NA-ACCORD. The values $(c_{1,17}, c_{1,18}, c_{1,19})$ were set to $(1.25, 50, -2000)$, and the values of $(c_{2,17}, c_{2,18}, c_{2,19})$ were set to $(-5, -200, 10000)$. Figure S1 in the Supplementary Material shows correlation plots for the covariates.
	
    Finally, the treatment indicator was generated as a function of a subset of covariates:
	\[A \sim \text{Bernoulli}(\text{logit}^{-1}(X_{13} - X_{14} - X_{15} - X_{1}))\]

	\subsection{Survival and censoring times} \label{sec:dgp}
	
\begin{figure}[!h]
    \centering
    \includegraphics[width=0.75\textwidth]{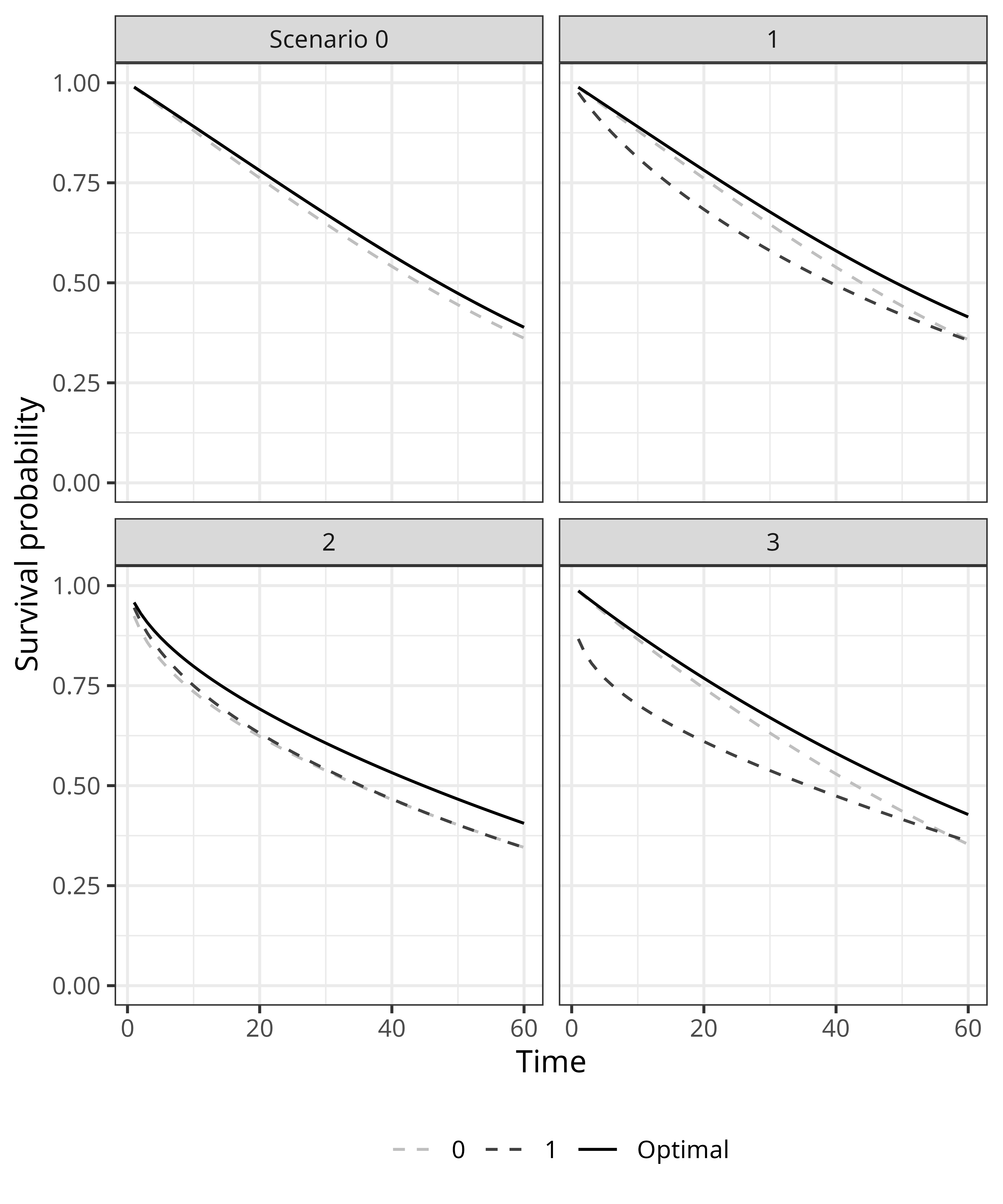}
    \caption{True survival probability curves under treatment 0 only, treatment 1 only, and optimal treatment rule, for simulation Scenarios 0-3.}
    \label{fig:true_surv}
    \end{figure}

\begin{figure}[!h]
     \centering
     \includegraphics[width=\textwidth]{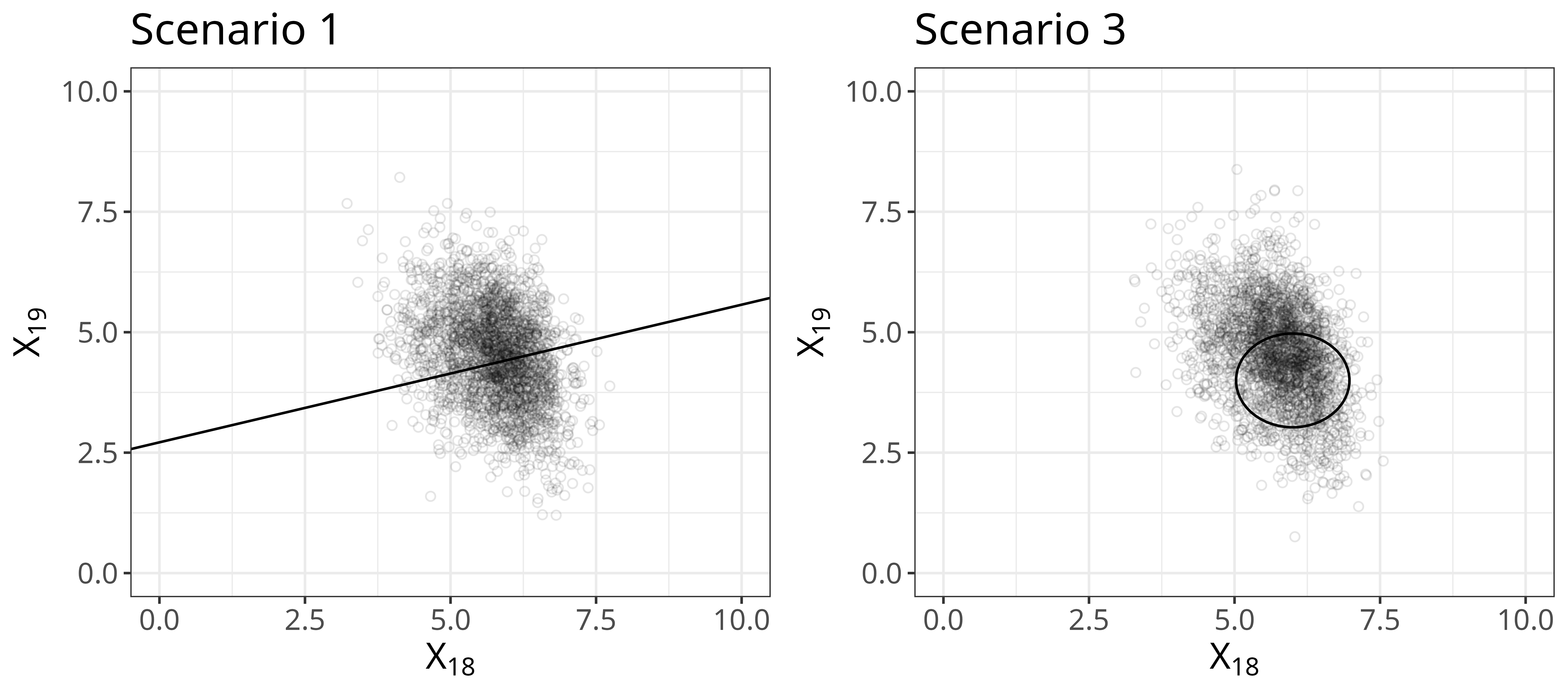}
    \caption{Visual representations of optimal treatment rules for simulation Scenarios 1 and 3. The optimal treatment rules for these scenarios are shown here since the rules depend on only two covariates. In Scenario 1, the optimal treatment rule assigns observations below the line to treatment 1. In Scenario 3, the optimal treatment rule assigns observations inside the circle to treatment 1.}
    \label{fig:treat_b}
    \end{figure}

	Survival and censoring times were generated from the Gompertz distribution: $C \sim  \text{Gompertz}(a,b \exp(f(X,A;\alpha)))$ and $T \sim \text{Gompertz}(\gamma, \lambda \exp(f(X,A;\beta)))$, where the parameters $(a,b,\gamma,\lambda) = (0.03, 0.01, 0.02, 0.002)$ were constant across simulations, $f(X,A;\alpha)$ and $f(X,A;\beta)$ were linear functions in $X$ and $A$, and $Q \sim \text{Gompertz}(\eta,\phi)$ indicates $Q$ has distribution function $P(Q \leq q) = 1 - \exp(-\phi / \eta [\exp(\eta q) - 1])$. Parameter values of the Gompertz distributions were selected to mimic the NA-ACCORD data where few individuals experienced the event at the beginning of the study and many were administratively right-censored. For all simulations, the censoring process used $f(X,A;\alpha) = -4 + 0.25X_{16} + 0.2X_{18}$. See Figure S2 in the Supplementary Material for a plot of the probability of the survival time being uncensored, i.e., $P(T_i < C_i)$, for 2,500 simulated individuals. Scenarios varied in the specification of $f(X,A;\beta)$ in the survival time generation. Letting $\beta = (\delta, \eta)$, the function $f(X,A;\beta)$ was defined to be 
    \begin{equation} \label{eq:time_gen}
        f(X,A;\delta,\eta) = g(X;\delta) + A \cdot h(X;\eta),
    \end{equation}
	where $g(X;\delta)$ represents the common dependence on covariates shared by the survival time processes for both treatment groups and $h(X;\eta)$ represents the extra contribution of the covariates on the survival time distribution when treatment $A=1$. The optimal treatment rule depends on $h$ directly; that is, $d^{opt}(X)=I(h(X;\eta)<0)$. Four different survival distributions were considered, shown in Table \ref{tab:tab2}. Figure \ref{fig:true_surv} shows the survival curves under these four scenarios. Since the optimal treatment rules for Scenarios 1 and 3 depend on only two covariates, the treatment rule boundaries are plotted and shown in Figure \ref{fig:treat_b}. 
 
	For all scenarios the average treatment effect of treatment 1 relative to treatment 0 is approximately 0, similar to the findings of a recent analysis of the NA-ACCORD \cite{lu_clinical_2021}. This simulation setting represents the case where two treatments show similar average effectiveness but where there is an advantage gained by recommending different treatments to different subgroups. In particular, the regret of both static rules, $R(1)$ and $R(0)$, were set to about 6 percent for all simulations.
	
	\subsection{Performance metrics}
	
	Three performance metrics were considered related to estimation of the value function under the optimal treatment rule, $V(d^\text{opt})$, for generic estimators $\widehat{V}$ and $\widehat{d}^\text{opt}$ of $V$ and $d^\text{opt}$, respectively. Performance metric (i) is $\widehat{V}(d^\text{opt})-V(d^\text{opt})$, which quantifies how well an estimator for $V$ performs and ignores variation from estimating $d^\text{opt}$. Performance metric (ii) is $V(\widehat{d}^\text{opt})-V(d^\text{opt})$, i.e., the difference in the true value of the estimated and optimal treatment rules. This quantity isolates the performance of optimal treatment rule estimators by ignoring estimation of the value function. Lastly, performance metric (iii) is $\widehat{V}(\widehat{d}^\text{opt})-V(d^\text{opt})$, i.e., the difference between the value estimate one could compute in a real study compared to the true value under the optimal treatment rule. 
	
	Of secondary importance in evaluating the performance of estimators for the optimal treatment rule is the mis-classification rate: $n^{-1} \sum_{i=1}^{n} I(\widehat{d}^\text{opt}(X_i) \neq d^\text{opt}(X_i))$. Since the goal of estimating optimal treatment rules is to maximize the value function, the mis-classification rate is less important than the other metrics because it is possible for an estimated optimal treatment rule to have a higher mis-classification rate while also having a higher value than some other estimated optimal treatment rule.
	
	Performance metric (i) was used to select the best estimator of $V$ out of the three defined above. Specifically, for 1,000 simulations under Scenario 2, performance metric (i) was computed using each of the three value function estimators and the mean and standard deviations across simulations (i.e., the empirical bias and empirical standard error of the value estimators) were compared. The best value estimator was then exclusively used in computing performance metric (iii).

    \subsection{Implementation}

    All estimators of the value function considered rely on estimation of the propensity score $\pi(X_i)$ and the probability of being uncensored $P(C_i \geq s|X_i, A_i)$. In these simulations, logistic regression and the Cox proportional hazards model with the Breslow estimator were used to estimate $\pi(X_i)$ and $P(C_i \geq s|X_i, A_i)$, respectively. In estimating $P(C_i \geq s|X_i, A_i)$, the probability of remaining not lost to follow-up was modeled, where time to loss to follow-up was subject to right-censoring due to $T_i$ or end-of-study. When an estimator required a parameterization of the treatment rule, a linear treatment rule was chosen in all cases except for OWL where both linear and radial kernels were implemented. Causal survival forests, Bayesian additive regression trees, and OWL with a radial kernel were the only methods that allowed for nonlinear treatment rules. All sample splitting estimators used five folds.

    R (version 4.1.0) \cite{r_core_team_r_2021} was used for this study. The \texttt{tidyverse} package \cite{wickham_welcome_2019} was used for general data manipulation, and the \texttt{survival} package \cite{therneau_package_2023} was used for fitting Cox proportional hazard models. The \texttt{glmnet} package \cite{tay_elastic_2023} was used for fitting penalized models, \texttt{grf} \cite{tibshirani_grf_2021} for causal survival forests, \texttt{BART} \cite{sparapani_nonparametric_2021-1} for Bayesian additive regression trees, \texttt{rgenoud} \cite{sekhon_genetic_1998} for the genetic algorithm, \texttt{WeightSVM} \cite{xu_weightsvm_2021} for outcome weighted learning, and \texttt{DynTxRegime} \cite{holloway_dyntxregime_2020} for residual weighted learning. Code to reproduce the results of this study is available; see the Data Availability Statement for details.

	\section{Simulation results}
	
	When evaluating performance metric (i), all value estimators performed well and had zero empirical bias to three decimal places. The IPW-KM ($\widehat{V}_1$) estimator had slightly smaller empirical standard error (0.014) compared to the IPW-HT and IPW-HJ estimators (0.016 and 0.015, respectively). Figure S3 in the Supplementary Material shows boxplots of $\widehat{V}_j(d^\text{opt}) - V(d^\text{opt})$ for $j=1,2,3$ under Scenario 2. Given the slightly better performance, the IPW-KM estimator was chosen as the value estimator in subsequent evaluations.

    Supplementary Figures S4 and S5 show boxplots of performance metric (ii) for each optimal treatment rule estimator and scenario considered using the whole sample and sample splitting, respectively. The performances were similar when either the whole sample or sample splitting was used for optimal treatment rule estimation, an unsurprising result since the value function is assumed known. Otherwise the optimal treatment rule estimators performed similarly relative to each other as in performance metric (iii), discussed below.
    
\begin{figure}[!h]
    \centering
    \includegraphics[width=\textwidth]{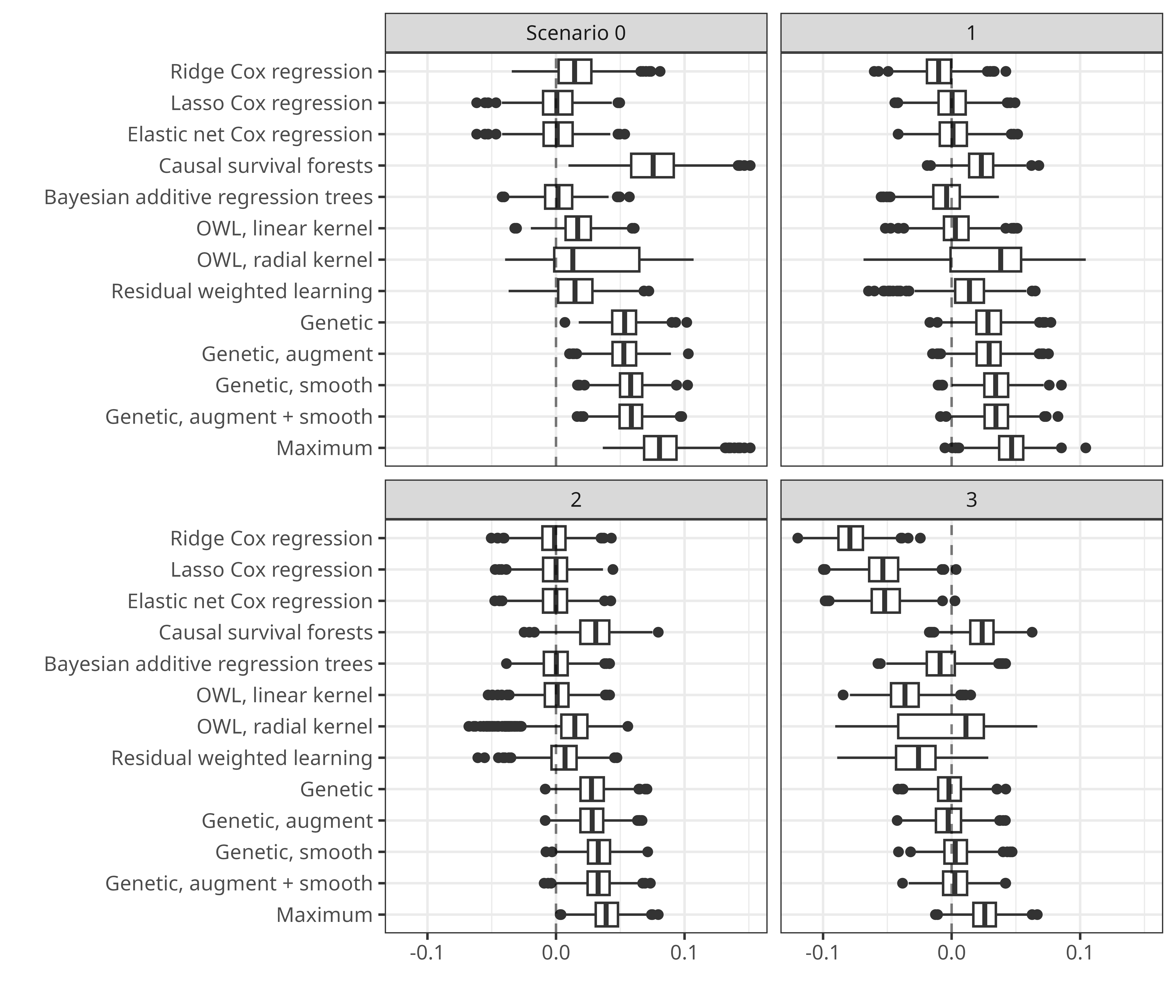}
    \caption{Boxplots of $\widehat{V}(\widehat{d}^\text{whole})-V(d^\text{opt})$ (performance metric (iii)), the difference in estimated survival probability under whole sample estimated optimal treatment rules and survival probability under the optimal treatment rule, at time 60, for simulation Scenarios 0-3, 500 simulations, $n=2500$.}
    \label{fig:metric3_whole}
    \end{figure}

\begin{figure}[!h]
    \centering
    \includegraphics[width=\textwidth]{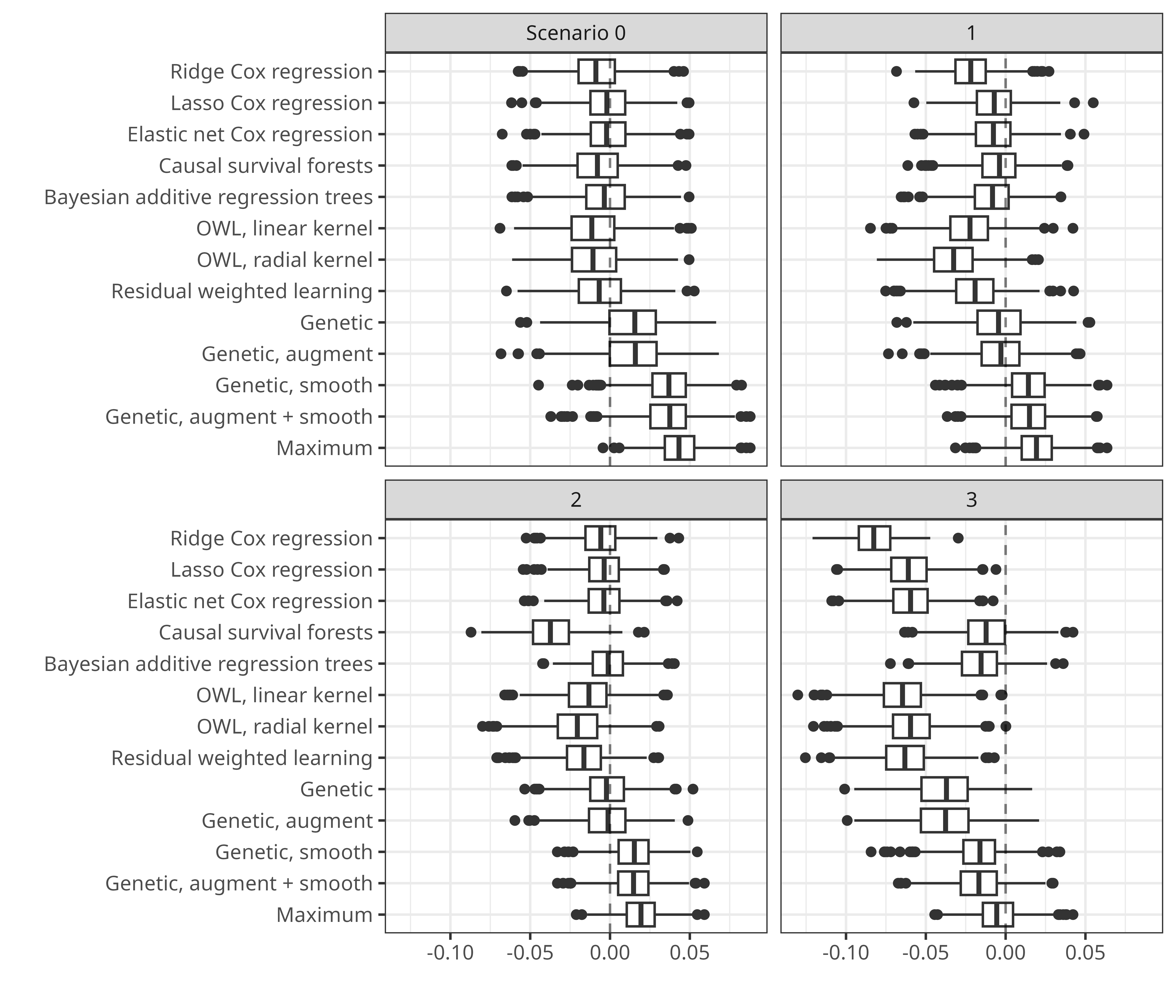}
    \caption{Boxplots of $\widehat{V}(\widehat{d}^{CV})-V(d^\text{opt})$ (performance metric (iii)), the difference in estimated survival probability under sample split estimated optimal treatment rules and survival probability under the optimal treatment rule, at time 60, for simulation Scenarios 0-3, 500 simulations, $n=2500$.}
    \label{fig:metric3_cv}
    \end{figure}

	Figures \ref{fig:metric3_whole} and \ref{fig:metric3_cv} show boxplots of performance metric (iii) for Scenarios 0 to 3 when the whole sample and sample split optimal treatment rule estimators were used, respectively. Scenario 0 in Figures \ref{fig:metric3_whole} and \ref{fig:metric3_cv} shows the possibility of over-estimating survival probability when there is no treatment effect heterogeneity but an optimal treatment rule depending on covariates is estimated anyway. In Scenario 0, treatment 1 is slightly better than treatment 0 but effects are not dependent on individual covariates. Methods that allow for estimated coefficients of a treatment rule to be zero (e.g., via elastic net or Lasso penalties) performed well here along with the BART method. Cox regression with the ridge penalty did not perform as well since the regularization does not set estimated coefficients to exactly zero. The genetic algorithm methods performed particularly poorly. The results of Scenario 1, where most but not all coefficients of the optimal treatment rule were zero, were similar though the genetic algorithm methods without smoothing performed better under sample splitting. In Scenario 2, the optimal treatment rule was a function of all measured covariates. Cox regression with a ridge penalty and OWL with a linear kernel thus performed better. The methods that performed well under Scenario 1 again performed well. Scenario 3 contained a non-linear optimal treatment rule so most estimators performed poorly. Surprisingly, however, the whole sample genetic algorithm methods performed well. 

    In general, the max ensemble estimator without sample splitting was shown to overestimate the value function under all scenarios. Using sample splitting attenuated the max ensemble estimator's over-optimism but did not eliminate it in many scenarios. BART performed comparatively well across scenarios, performing among the best in Scenarios 0-2 while performing better in Scenario 3 compared to other methods that assume a linear optimal treatment rule. However, it should be noted that BART was the most computationally intensive of the considered methods. 

    For the four simulation scenarios, mis-classification rates for the estimated optimal treatment rules when the estimators used the whole sample and sample splitting are shown in Supplementary Figures S6 and S7, respectively. These results were consistent with the results from performance metrics (ii) and (iii).

	\section{Discussion}

    The simulations presented in this study represent a use case for precision medicine in large observational datasets with right-censored outcomes. In these simulations, the average treatment effect was approximately zero while there was treatment effect heterogeneity such that the effect of the optimal treatment rule was greater than that of either treatment alone (except in Scenario 0 where treatment 1 was slightly better and there was no treatment effect heterogeneity). When treatments are estimated to have similar average effectiveness, deciding between treatments may be supported by estimating an optimal treatment rule in an observational study, potentially improving patient outcomes. 
 
	Though BART was perhaps the most flexible, different estimators performed better under different scenarios, making it difficult to recommend a single method to implement universally. The max ensemble estimator was shown to lead to over-estimation problems, but the sample split estimation method described in Section \ref{sec:sample_split} may help partially ameliorate this issue.
    
    Though an observational study motivated the simulations in this study, these results are also relevant for randomized clinical trials (RCTs) where survival times for some participants are right-censored. Here, the treatment assignment mechanism was unknown and had to be estimated. In an RCT, the propensity score is known but the framework and methods presented here are otherwise similar. One drawback of using observational datasets compared to data from RCTs is the potential for bias due to unmeasured confounders. On the other hand, observational studies often have the benefit of more closely matching the target population under real-life conditions. In either case, the ability to estimate optimal treatment rules depends on measuring treatment effect modifiers. 
    
    Future work could explore more extensive simulation scenarios motivated by other applications. In the simulations presented here, the treatment assignment mechanism was held fixed across simulations and only the outcome-treatment relationship was allowed to vary; in future research it may be of interest to further explore mis-specification of the treatment rule, outcome model, and propensity score. Furthermore, there may be clinical interest in examining the finite sample properties of the above estimators for target parameters of survival outcomes other than survival probability, such as RMST. Optimal treatment rule estimation is a fast growing field and though the estimators considered in this study represent a reasonable portion of estimators in use, there are many more estimators being developed along with software packages. This simulation studies examined here can be used to inform the choice of value estimators and optimal treatment rule estimators for data with right-censored outcomes.

    \section*{Acknowledgements}
    We thank Chanhwa Lee for helpful comments.

    \section*{Funding information}
    This work was supported by the National Institutes of Health (NIH): grant numbers P30AI050410, R01AI157758, and T32ES007018. 

    \section*{Conflict of interest statement}
    The authors declare no potential conflict of interests.
	
	\newpage
	
	\section*{References}

    \bibliographystyle{wileyNJD-AMA.bst}
    {\def\section*#1{}
    \bibliography{main}
    }

    \newpage
    
    \section*{Tables}

    \begin{table}[h]
    \begin{tabular}{|l|l|l|l|}
    \hline
    \textbf{Covariate} & \textbf{Characteristic}                               & \textbf{NA-ACCORD} & \textbf{Simulation} \\
     & & \textbf{(n=15,993)} & \textbf{(n=2,500)} \\ \hline
    $A$ & InSTI treatment (\%)                                  & 36.4                          & 37.9                          \\ \hline
    $X_1$ & Injection drug use (\%)                               & 10                            & 12.4                          \\ \hline
    $X_2$ & Male-to-male sexual contact (\%)                      & 48.3                          & 43.2                          \\ \hline
    $X_3$ & Risky heterosexual behavior (\%)                      & 21.1                          & 20.9                          \\ \hline
    $X_4$ & Previous AIDS diagnosis (\%)                          & 7.6                           & 7.2                           \\ \hline
    $X_5$ & Hepatitis B (\%)                                      & 4                             & 5.4                           \\ \hline
    $X_6$ & Hepatitis C (\%)                                      & 10.5                          & 12.4                          \\ \hline
    $X_7$ & Previous depression diagnosis (\%)                    & 11.9                          & 12.6                          \\ \hline
    $X_8$ & Previous anxiety diagnosis (\%)                       & 8.9                           & 10                            \\ \hline
    $X_9$ & Diabetes mellitus (\%)                                & 5.2                           & 4.7                           \\ \hline
    $X_{10}$ & Hypertension (\%)                                     & 16.3                          & 17                            \\ \hline
    $X_{11}$ & Statin prescription (\%)                              & 7.2                           & 8.7                           \\ \hline
    $X_{12}$ & Elevated total cholesterol (\%)                       & 4.6                           & 5.6                           \\ \hline
    $X_{13}$ & Female (\%)                                           & 12.5                          & 13.1                          \\ \hline
    $X_{14}$ & Black race (\%)                                       & 43.5                          & 42.8                          \\ \hline
    $X_{15}$ & Hispanic ethnicity (\%)                               & 13                            & 13.8                          \\ \hline
    $X_{16}$ & Age, median (IQR), years                              & 40 (30-50)              & 40.2 (30.3-52.1)              \\ \hline
    $X_{17}$ & Body mass index, median (IQR)                         & 25.1 (22.3-28.6)              & 25.1 (21.4-29.1)              \\ \hline
    $X_{18}$ & Baseline CD4 T-cell count, median (IQR), cells/$\mu$L & 332 (177-485)           & 324 (206-472)                 \\ \hline
    $X_{19}$ & Baseline viral load (log10), median (IQR), copies/mL  & 4.6 (3.9-5.1)                 & 4.6 (3.9-5.3)                 \\ \hline
    \end{tabular}
    \caption{Summary of patient characteristics from the North American AIDS Cohort Collaboration on Research and Design (NA-ACCORD) \cite{lu_clinical_2021} and from a simulation dataset.}
    \label{tab:tab1}
    \end{table}

    \newpage

    \begin{table}[h] 
    \centering 
    \begin{tabular}{|c|c|c|} 
    \hline 
    \textbf{Scenario} & \textbf{$g(X)$} & \textbf{$h(X)$} \\ \hline 
    0 & $-1-2X_{3}+0.1X_{18}+0.5X_{19}$ & $-0.1$ \\ \hline 
    1 & $-1-2X_{3}+0.1X_{18}+0.5X_{19}$ & $-1.9-0.2X_{18}+0.7X_{19}$ \\ \hline 
    2 & $-1.5-0.2X_{16}+0.2X_{17}-0.2X_{18}$ & $-0.1+0.1X_{16}-0.1X_{17}+0.1X_{18}-0.1X_{19}$ \\
     & $+0.5X_{19}+1.1\sum\limits_{j=1}^{15}X_j$ & $+0.3\sum\limits_{j=1}^{6}X_j+0.3\sum\limits_{j=13}^{15}X_j-1.2\sum\limits_{j=7}^{12}X_j$ \\ \hline
    3 & $-0.4+0.3X_{18}+0.8X_{19}$ & $-0.55+(X_{18}-6)^2-(X_{19}-4)^2$ \\ \hline 
    \end{tabular} 
    \caption{Survival times were generated as $\text{Gompertz}(0.02, 0.002 \exp(g(X) + A \cdot h(X))$; see Section \ref{sec:dgp} for details.}
    \label{tab:tab2} 
    \end{table}

    \newpage
    \section*{Figures}

    \newpage
    
    \newpage

    \newpage
    
\end{document}